\documentclass[useAMS,usenatbib]{mnras}

\usepackage{placeins}
\usepackage{graphicx}
\usepackage{amsmath}
\usepackage{rotating}
\usepackage{amssymb,amsfonts,hyperref,color}

\newcommand\be{\begin{equation}}
\newcommand\en{\end{equation}}

\newcommand\etal{{\rm et al}.\ }

\title[Massive lopsided transition discs I]{Gas and dust hydrodynamical simulations of massive lopsided transition discs -- I.  Gas distribution}

\author[Zhu and Baruteau]{Zhaohuan Zhu$^{1,3}$\thanks{E-mail: zhzhu@astro.princeton.edu(ZZ);
Clement.Baruteau@irap.omp.eu(CB)} and Cl{\'e}ment Baruteau$^{2}$\\
$^{1}$Department of Astrophysical Sciences, 4 Ivy Lane, Peyton Hall,
Princeton University, Princeton, NJ 08544, USA\\
$^{2}$CNRS / Institut de Recherche en Astrophysique et Plan{\'e}tologie, 14 avenue Edouard Belin, 31400 Toulouse, France\\
$^{3}$Hubble Fellow.}

\date{Accepted 2016 January 21. Received 2016 January 21; in original form 2015 August 4}
 
\pagerange{\pageref{firstpage}--\pageref{lastpage}} \pubyear{2015}

\def\LaTeX{L\kern-.36em\raise.3ex\hbox{a}\kern-.15em
    T\kern-.1667em\lower.7ex\hbox{E}\kern-.125emX}

\begin{document}

\label{firstpage}

\maketitle

\begin{abstract}
  Motivated by lopsided structures observed in some massive transition
  discs, we have carried out 2D numerical simulations to
  study vortex structure in massive discs, including the effects of
  disc self-gravity and the indirect force which is due to the
  displacement of the central star from the barycentre of the system
  by the lopsided structure.  When only the indirect force is
  included, we confirm the finding by \cite{MC15} that the vortex
  becomes stronger and can be more than two pressure scale heights
  wide, as long as the disc-to-star mass ratio is $\gtrsim$1$\%$. Such
  wide vortices can excite strong density waves in the disc and
  therefore migrate inwards rapidly.  However, when disc self-gravity
  is also considered in simulations, self-gravity plays a more
  prominent role on the vortex structure.  We confirm that when the
  disc Toomre Q parameter is smaller than $\pi/(2h)$, where $h$ is the
  disc's aspect ratio, the vortices are significantly weakened and
  their inward migration slows down dramatically.  Most importantly,
  when the disc is massive enough (e.g. Q$\sim$3), we find that the
  lopsided gas structure orbits around the star at a speed
  significantly slower than the local Keplerian speed.   This
    sub-Keplerian pattern speed can lead to the concentration of dust
    particles at a radius beyond the lopsided gas structure (as shown
    in Paper II).  Overall, disc self-gravity regulates the vortex
  structure in massive discs and the radial shift between the gas and
  dust distributions in vortices within massive discs may be probed by
  future observations.
\end{abstract}


\section{Introduction}
{ Transition discs are protoplanetary discs whose inner regions have undergone substantial clearing  
 \citep[see the review by][]{Espaillat14}. } 
Recent submm interferometric observations have suggested significant
non-axisymmetric features in these discs (HD 142527 from
\citealt{Casassus13}; Oph IRS 48 from \citealt{vanderMarel13};
LkH$\alpha$ 330 from \citealt{Isella13}). In the extreme case of Oph
IRS 48, there is a highly asymmetric crescent-shaped dust structure
between 45 and 80 AU from the star.  The peak emission from this dust
structure is at least 130 times stronger than the upper limit of the
opposite side of the disc. Observations at longer wavelengths
(e.g. 8.8 mm with Australia Telescope Compact Array for HD 142527,
\citealt{Casassus15}) reveal an even more compact structure.  These
observations suggest that some dust trapping mechanism is operating in
the azimuthal direction of the disc.

{ Theoretically, it has been known that anticyclonic vortices can be
  long lived \citep{godon2000} and can efficiently trap dust particles
  \citep{barge1995,adams1995,tanga1996,chavanis2000,lyra2009a,Johansen2004,meheut2012,zhu2014}.}
 Using the vortex gas structure derived by \cite{Kida} and
  \cite{goodman1987}, particle distribution within the vortex has been
  derived and compared with observed asymmetric disc structures
  \citep{lyra2013}.  By constructing realistic three-dimensional MHD
simulations including dust particles, \cite{zhustone2014} have found
that dust trapping vortices can reproduce ALMA observations reasonably
well.

However, some transition discs are quite massive. For an example, the
disc-to-star mass ratio is $\sim 4\%$ in the HD 142527 system.
Recently, \cite{MC15} has suggested a different mechanism for
azimuthal particle trapping in massive discs by allowing the star to
move around the barycentre of the system due to the gravitational pull
of the massive lopsided structure.  They suggest that, if the
azimuthal structure in these discs is massive enough, its
gravitational force to the central star can displace the star from the
barycentre of the system. Such displacement causes an $m=1$ mode
indirect force to the disc, which can lead to an asymmetric horseshoe
flow pattern. This pattern can be self-sustaining as long as it can
lead to enough gravitational force required to offset the
star. However, \cite{MC15} applied a prescribed indirect force to the
disc, and it is unclear if this mode can sustain with the indirect
force self-consistently calculated from the disc asymmetric structure
itself.

{ Furthermore, the disc self-gravity has been ignored in
  \cite{MC15}. If the disc is so massive that the indirect force can
  become important, the disc self-gravity should play an even more
  important role in shaping the asymmetric structure.}  Suppose the
lopsided pattern at a distance $r$ from the star has a mass 
$m_{LP}$, the acceleration of the central star is on the order of
$\sim Gm_{LP}/r^2$ and thus the indirect force felt by unit mass in
the disc is $\sim Gm_{LP}/r^2$. On the other hand, the force due to
the self-gravity of the asymmetric pattern is on the order of
$Gm_{LP}/\Delta r^2$ where $\Delta r$ is the scale of the asymmetric
pattern.  Since $\Delta r\lesssim$ r, the disc self-gravitating force should be
larger than the indirect force.  Previous simulations by
\cite{LinMK12} have shown that self-gravity can suppress large scale
vortices produced by Rossby Wave Instability (RWI)
\citep{lovelace1999}.  Linear instability analysis by
\cite{lovelace2013} has also suggested that disc self-gravity can
stabilize the $m=1$ RWI mode when the disc's Toomre Q parameter
$Q<(\pi/2)(r/H)$ where $H$ is the disc's scale height.  On the other
hand, the indirect force is a large scale $m=1$ driving force, which
can facilitate $m$=1 large scale structure.  Thus, a self-consistent
model including both the indirect force and disc self-gravity is
necessary.

{ In this paper, we present a self-consistent model including both the
  indirect force and disc self-gravity, focusing on their effects on
  the gas structure of the vortex.} In the subsequent paper 
 \citep{BZ15}, we will study how these effects impact particle
concentration in vortices.  In \S 2, we first follow \cite{MC15} using
test particle method to study asymmetric structures in pressureless
non-selfgravitating discs. Then we introduce our hydrodynamical
simulations including both disc pressure and self-gravity in \S
3. Results are presented in \S 4 and summarized in \S 5.

\section{``Fast modes'' in pressureless non-selfgravitating fluids} {
  Although the horseshoe solution led by the indirect force
  \citep{MC15} seems to be distinct from the traditional vortex
  solution \citep*{goodman1987}, both solutions are steady in a
  Keplerian rotating reference frame.} Thus, they are both classified
as ``fast modes'', in contrast with the ``slow modes''
\citep{tremaine2001, Lin2015} that is almost steady in an inertial
reference frame.   In this section, we present the traditional
  vortex solution and the new horseshoe solution together in the
  pressureless fluid.

In a frame which is centered on the star and rotates at the angular
frequency $\mathbf{\Omega}_{f}$, the Euler equation for the disc
becomes
\begin{eqnarray}
  \frac{\partial \mathbf{v}}{\partial t}+(\mathbf{v}\cdot \nabla)\mathbf{v}&=&-\frac{1}{\rho}\nabla P-\nabla \Phi_{*} -\nabla \Phi_{ind}- \nabla \Phi_{sg} \nonumber\\
  &&-\mathbf{\Omega}_{f}\times(\mathbf{\Omega}_{f}\times \mathbf{r})-2\mathbf{\Omega}_{f}\times \mathbf{v}\,,\label{eq:euler}
\end{eqnarray}
where $\Phi_{*}$, $\Phi_{ind}$ and $\Phi_{sg}$ are the potential due
to the direct gravitational force from the star, the indirect force
due to the acceleration of the reference frame (or the star), and the
disc self-gravity. Specifically
$\Phi_{ind}=-\mathbf{a_{c}}\cdot\mathbf{r}$, where $\mathbf{a_{c}}$ is
the acceleration of the central star.  The last two terms in Equation
\ref{eq:euler} are the centrifugal force and the Coriolis force. The
pressure term, indirect force term and self-gravity term are all
related to the disc density distribution ($\rho$) which has to be
solved with the continuity equation.

 For a ``fast mode'' which is steady in the rotating frame, we
  have $\partial \mathbf{v}/\partial t$=0 so that the fluid
  streamlines and trajectories coincide. We can then use the
trajectories of test particles to study the flow streamlines. Such
test particle method has been widely used in studying discs in binary
systems \citep{paczynski1977} and galactic dynamics
\citep{contopoulos1979, binney1991}.  The position and velocity of the
test particle ($\mathbf{r}$, $\mathbf{u}$) are
\begin{eqnarray}
  \frac{d\mathbf{r}}{dt}&=&\mathbf{u}\label{eq:stra}\\
  \frac{d\mathbf{u}}{dt}&=&-\frac{1}{\rho}\nabla P-\nabla \Phi_{*} -\nabla \Phi_{ind} - \nabla \Phi_{sg} \nonumber\\
  &&-\mathbf{\Omega}_{f}\times(\mathbf{\Omega}_{f}\times \mathbf{r})-2\mathbf{\Omega}_{f}\times \mathbf{u}\,. \label{eq:utra}
\end{eqnarray}
If the indirect force term is known beforehand and the density related
terms (pressure, and self-gravity) are small compared to other terms,
we can ignore these density related terms to simplify the problem
further and search all possible ``fast modes'' as in \cite{MC15}.
Equations \ref{eq:stra} and \ref{eq:utra} can be solved given initial
position and velocity for a test particle.  Although we can have
infinite trajectories, for a given flow only trajectories that do not
intercept each other or itself represent allowed streamlines. For a
steady flow, the streamlines also have to be closed, unless there is
external mass flowing in. Thus, in the rest of this section, we will
look for closed non-intercepting trajectories through numerical
integration of Equations \ref{eq:stra} and \ref{eq:utra} using the
fixed timestep 5th order Runge-Kutta method.
 
To search these possible ``fast mode'' flow patterns, we choose
$\Omega_{p}=\Omega_{f}=1$ where $\Omega_{p}$ is the pattern speed. In
this case, the pattern is steady in the rotating frame.  The lopsided
pattern is assumed to be symmetric to the $x$ axis so that the
indirect force is along the $x$ axis.  The indirect force is opposite
to the direction of the acceleration of the central star, and it is a
constant force anywhere in the disc.  The prescribed indirect force is
determined from hydrodynamical simulations in \S 3.  In the unit that
$GM_{*}$=1, the indirect force is $-5.625\times 10^{-3}\hat{x}$ (\S
3). Test particles are launched at the $x$ axis with $v_{x}=0$.  Since
the force is symmetric to the $x$ axis and the particle is initialized
with only $v_{y}$ at the $x$ axis, this symmetry implies that the
particle trajectory is a non-intercepting closed orbit only if
$v_{x}=0$ when the particle crosses the $x$ axis for the first time
after it has been launched.  By varying $v_{y}$, we search these
closed non-intercepting orbits.

\begin{figure}
\centering
\includegraphics[trim=2.3cm 0.0cm 3.cm 0.cm, width=0.23\textwidth]{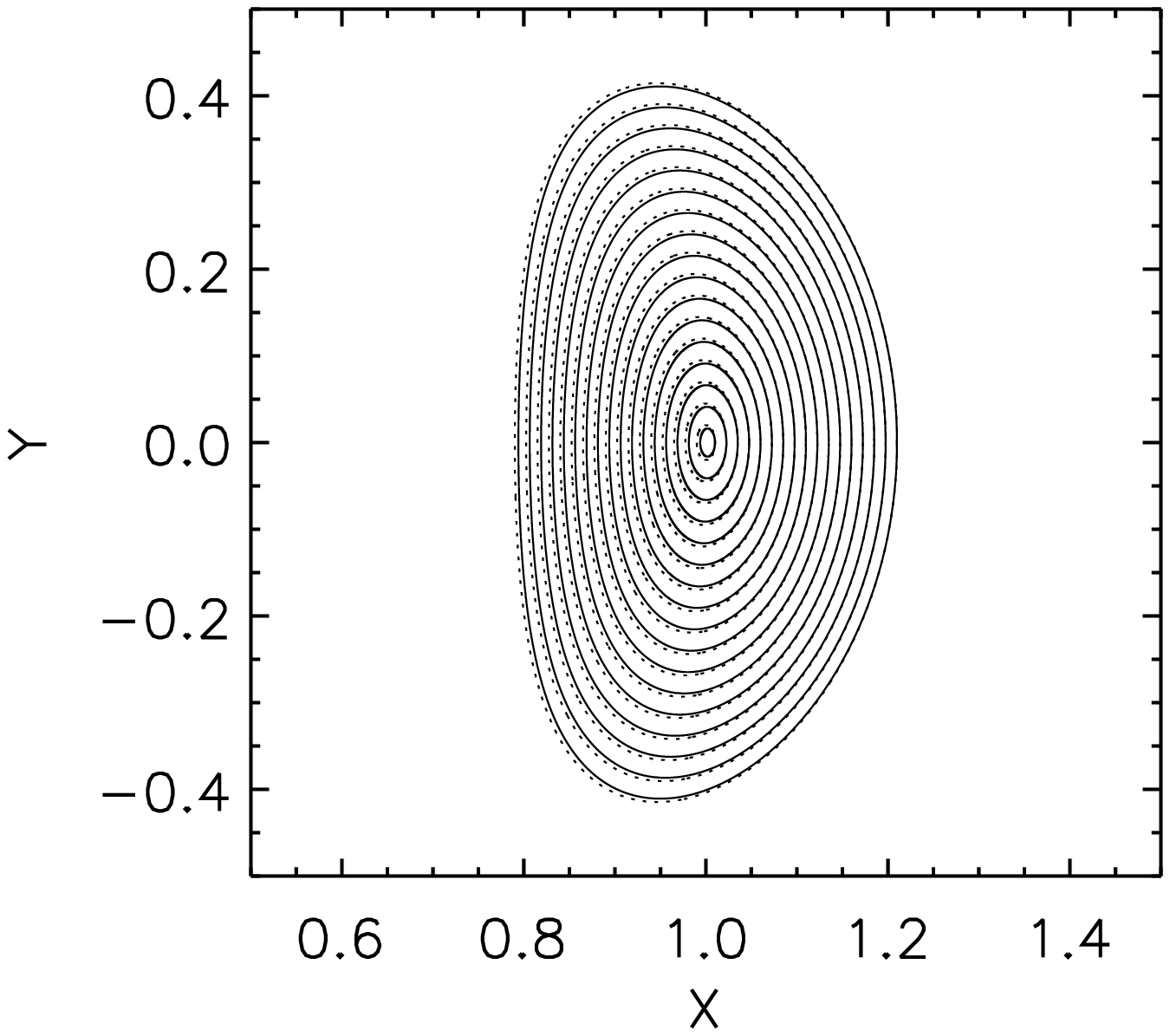} 
\includegraphics[trim=2.3cm 0.0cm 3.cm 0.cm, width=0.23\textwidth]{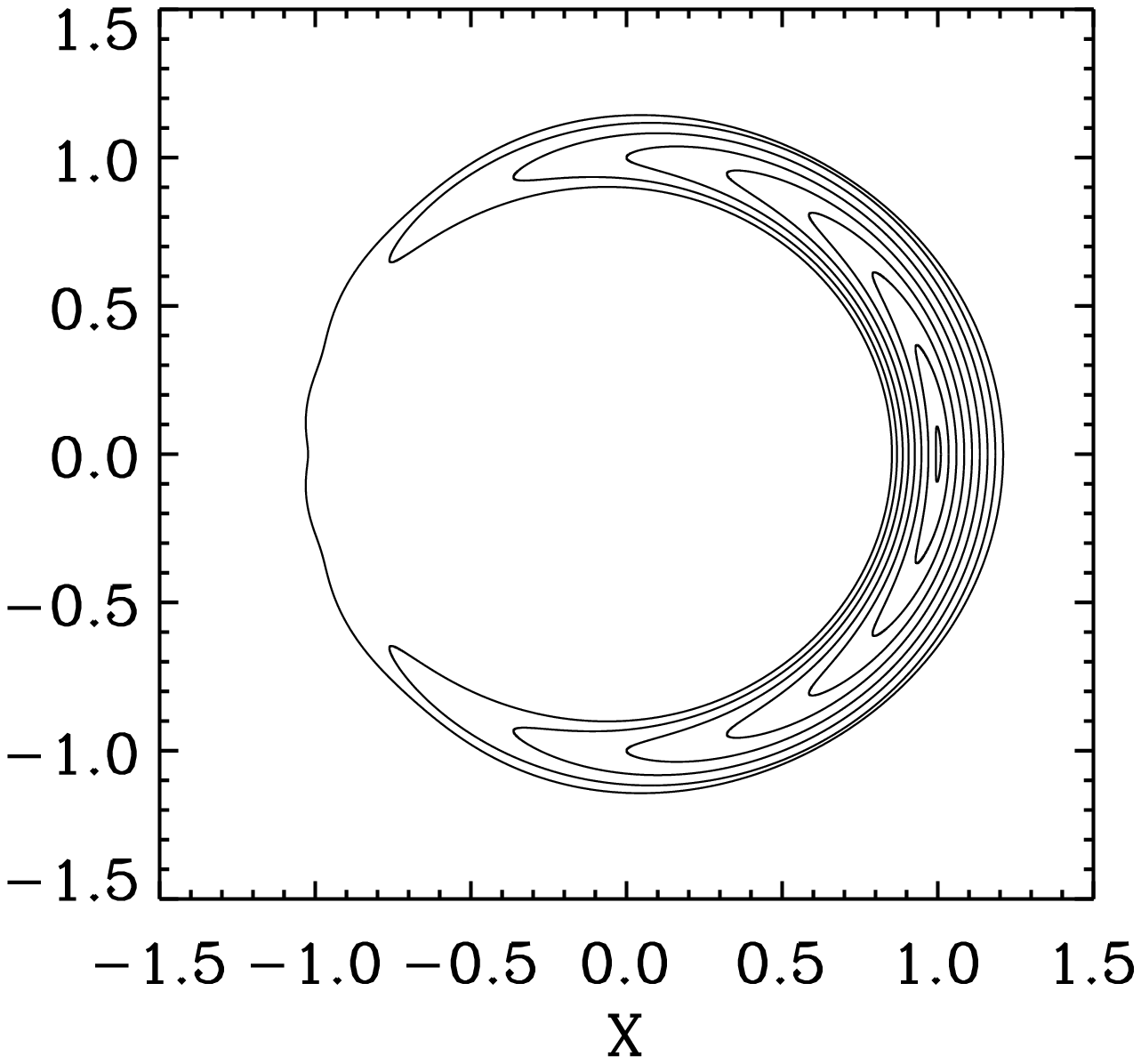}
\vspace{-0.3 cm}
\caption{Two closed non-crossing orbit configurations when the
  center-of-mass is not at the star. Left panel: the eccentric orbits
  forms a vortex pattern with an aspect ratio of 2. The solid curves
  have considered the indirect force while the dotted curves have not
  considered the indirect force. Clearly, the indirect force has small
  effect on these eccentric orbits.  Right panel: similar to Mittal \&
  Chiang (2015) figure 1, horseshoe shaped orbits form due to the
  indirect force.  In both panels, the orbits are initiated at the
  same $x$ positions.  The lopsided pattern in the right panel is much
  more elongated than the one in the left panel.
} \label{fig:indirect}
\end{figure}

We have searched closed non-intercepting orbits of particles with and
without including indirect force due to the large scale lopsided
pattern.  Without the indirect force, the most simple ``fast mode'' is
the traditional pressureless elliptical vortex having an aspect
ratio($\chi$) of 2. This is shown in the left panel of Figure
\ref{fig:indirect}.  Fluid streamlines in this mode are the
trajectories of test particles undergoing epicyclic motion in the
rotating frame.  The guiding center approximation suggests that, in
the rotating frame whose angular frequency ($\Omega_{f}$) equals the
test particles' mean motion, eccentric particles undergo anticyclonic
epicyclic motions and form ellipses with aspect ratios of 2. All test
particles in the left panel of Figure \ref{fig:indirect} have their
mean motion equal to $\Omega_{f}$ but different eccentricities.  This
``fast mode'' is the only ``fast mode'' when there is no indirect
force.  \footnote{In a frame rotating at $\Omega_{f}$, an eccentric
  particle must have its mean motion equal to $\Omega_{f}$ or in
  first-order mean motion resonance with $\Omega_{f}$ to form a
  non-intercepting closed orbit.  However, streamlines cannot consist
  of different first-order mean motion resonant orbits (e.g.  2:1,
  3:2, 4:3, ...)  since these orbits are discretized in space while
  streamlines should be continuous in space.  Streamlines can neither
  consist of orbits in the same first-order mean motion resonance
  (e.g.  2:1) but having different eccentricities, since these orbits
  will intercept each other.  Thus, the only possible fluid
  streamlines consist of trajectories of test particles having their
  mean motion equal to $\Omega_{f}$ but different eccentricities, as
  shown in the left panel of Figure \ref{fig:indirect}.}  This
  mode can also be derived from the traditional compressible vortex
  solution under the pressureless limit (\S 3).  

When the indirect force due to the displacement of the star from the
barycentre of the system is included, two classes of closed
non-intercepting smooth orbits with dramatically different velocities
have been found. The first class of orbits forms the $\chi\sim2$
vortex solution above, with a slight modulation by the indirect
force. In this mode, the test particles only need one orbital time
($2\pi/\Omega_{f}$) to finish one circle around the vortex center, and
the acceleration due to the addition of the star's gravity and the
inertial force is on the order of $\sim eGM/r^{2}$, much larger than
the indirect force. Thus, the indirect force has very little effect on
this class of orbits (red curves in Figure \ref{fig:indirect}).  The
aspect ratios of the ellipses are still $\sim$ 2.

Another class of closed non-intercepting orbits in the rotating frame
is from slowly moving test particles (the right panel of Figure
\ref{fig:indirect})\footnote{ Starting from some initial positions, we
  can find multiple smooth closed orbits while at some positions only
  roughly smooth closed orbits can be found.}, as originally shown by
\cite{MC15}. Different from the vortex solution above where test
particles undergo rapid epicyclic motion, test particles in this mode
undergo slow horseshoe orbits induced by the indirect force. The
libration frequency for the particle is around
$\sqrt{3\mu}\Omega_{f}$\citep{MC15}, where $\mu$ is the ratio between
the distance from the barycentre to the star and is equal to 0.0056 in
our setup.  Test particles in the right panel of Figure
\ref{fig:indirect} thus need $\sim 10$ orbital time to finish one
circle around the vortex center.  The motion of these test particles
is so slow in the rotating frame that inertial forces almost balance
the star's gravity, and the indirect force plays a crucial role in
determining the orbits.  During one horseshoe orbit, the test particle
loses and gains angular momentum from the indirect force.  In the
limit that the indirect force becomes zero, these horseshoe orbits
gradually become circular orbits around the central star.

\section{Hydrodynamical Simulations}
 Although the two ``fast modes'' in pressureless fluids have
  distinct aspect ratios, they become less distinct when the gas
  pressure is considered.  The compressible vortex solution by
  \cite{goodman1987} (GNG solution) suggests that, even without either
  the indirect force or disc self-gravity, the gas pressure alone can
  lead to vortices with any aspect ratio larger than 2.  In the
  special case that the vortex aspect ratio ($\chi$) equals 2, the GNG
  solution becomes the pressureless vortex solution with streamlines
  as those shown by the test particle method in the left panel of
  Figure \ref{fig:indirect}.  For a vortex with $\chi$ larger than 2,
  the pressure force starts to balance the tidal force, Coriolis force
  and centrifugal force. When $\chi$ becomes larger, the net force
  which leads to vortex rotation becomes smaller, so that the vortex
  rotates slower and is more elongated.  Eventually when $\chi$ keeps
  increasing, this net force can be so small that any additional force
  (such as the indirect force and self-gravity) starts to affect the
  vortex structure.

To study how the indirect force and disc self-gravity affect
  the vortex solution in compressible fluids, we have carried out two
  dimensional hydrodynamical simulations including both of these
  forces.
\subsection{Set-up}
{  The hydrodynamical code we used is 
FARGO-ADSG \citep{bm08a,bm08b} which is built on FARGO \citep{fargo1} but with optional self-gravity and energy equation.}
Since the numerical grid is centered at the central star, the velocity
equations in a rotating frame are
\begin{eqnarray}
\frac{\partial v_{r}}{\partial t}+v_{r}\frac{\partial v_{r}}{\partial r}+v_{\phi}\frac{\partial v_{r}}{r\partial \phi}-\frac{v_{\phi}^2}{r}=\nonumber\\
-\frac{\partial P}{\rho \partial r}-\frac{\partial \Phi_{*}}{\partial r}-\frac{\partial \Phi_{ind}}{\partial r}-\frac{\partial \Phi_{sg}}{\partial r}+\Omega_{f}^{2}r+2\Omega_{f} v_{\phi}\\
\frac{\partial v_{\phi}}{\partial t}+v_{r}\frac{\partial v_{\phi}}{\partial r}+v_{\phi}\frac{\partial v_{\phi}}{r\partial \phi}-\frac{v_{\phi}v_{r}}{r}=\nonumber\\
-\frac{1}{\rho r}\frac{\partial P}{\partial \phi}-\frac{1}{r}\frac{\partial \Phi_{*}}{\partial \phi}-\frac{1}{r}\frac{\partial \Phi_{ind}}{\partial \phi}-\frac{\partial \Phi_{sg}}{\partial \phi}-2\Omega_{f} v_{r}
\end{eqnarray}
where the indirect potential
$\Phi_{ind}=-\mathbf{a_{c}}\cdot\mathbf{r}$ and $\mathbf{a_{c}}$ is
the acceleration of the central star due to the disc's gravity. It is
calculated by integrating the gravitational acceleration from each
grid cell onto the central star.  Indirect forces are calculated at
each time step and added to the equation of motion.  The simulations
are run in the inertial frame centered at the star, but analyses shown
later are done in the rotating frame with $\Omega_{f}$ equal to the
pattern speed of the asymmetric disc structure so that the time
derivatives in these equations become almost zero.

To make each grid cell have the same length in both the radial and
azimuthal directions, our grids are uniformly spaced in ${\rm log}\,
r$ from $r_{in}$ to $r_{out}$ where $r_{in}=0.1 r_{0}$ and $r_{out}=10
r_{0}$.  Our standard simulations have 752 grids in the radial
direction, and 1024 grids in the azimuthal direction. When disc
self-gravity is included, a softening length of 0.3 disc scale height
is used in the self-gravity potential to mimic the effect of a finite
disc thickness. Outflow boundary condition is used at both inner and
outer boundaries.

To generate a vortex in the simulation, we initialize the disc surface
density with a density bump in the radial direction which will later
break into vortices through Papaloizou-Pringle instability
\citep{papaloizou1984,papaloizou1985} or Rossby Wave instability (RWI)
\citep{lovelace1999,li2000,li2001}. The initial density profile is
\begin{equation}
  \Sigma(r)=\Sigma_{0}\times\left\{\zeta+{\rm exp}\left[-\frac{(r-r_{0})^2}{2\sigma^2}\right]\right\}\,,\label{eq:sigma0}
\end{equation}
where $\zeta=0.01$, and $r_{0}=1$. 
\cite{lyra2009b} and \cite{Regaly2012} have showed that RWI can be excited 
in $\alpha$ disks only if the density jump is sharp enough with
 jump width less than about 2 pressure scale-height ($H$). 
 $H$ is defined as $c_{s}/\Omega_{K}$ where
$c_{s}$ is the disc sound speed, and $\Omega_{K}$ is the orbital
frequency. Thus, we choose   
$\sigma=2 H(r_{0})$. To trigger $m=1$ mode in the instability, we add a small
perturbation to the gas surface density as in \cite{HPS92}
\begin{equation}
\Sigma(r,\phi)=\Sigma(r)\times\left[1+10^{-3}\times {\rm cos}(\phi)\times {\rm sin}\left(\pi \frac{r-r_{in}}{r_{out}-r_{in}}\right)\right]\,.
\end{equation}
We have run the simulations for 500 orbits, and in this paper we refer
one orbit as the orbital time ($2\pi/\Omega$) at $r_{0}$.

We use the isothermal equation of state so that $P=\Sigma c_{s}^2$.
The choice of an isothermal equation of state is for the sake of
simplicity, but we note that the inclusion of an energy equation
impacts the growth and saturation phases of the RWI \citep[see
e.g.,][where the RWI is induced by a gap-opening planet]{LesLin15}.
Since the whole disc has the same temperature, the disc aspect ratio
$h\equiv H/r$ increases as $\sqrt{r}$. We take $h=0.1$ at $r=r_{0}$,
which is typical for a protoplanetary disc at 10s of AU.  A small
viscosity with $\alpha=10^{-6}$ is included in the simulation. We find
that adding this small viscosity significantly improves the
convergence of the simulation.  To explore the effect of $\alpha$ on
the vortex structure, we have also increased $\alpha$ to 10$^{-3}$ for
our fiducial run.

To study the effect of indirect force and self-gravity on the vortex
structure, we choose discs which are relatively massive but still
gravitationally stable.  Our fiducial run assumes
$\Sigma_{0}$=$5\times10^{-3}$ (models g5 in Table 1) which has a total
mass of 0.017 $M_{*}$, and the Toomre parameter, $Q\equiv
c_{s}\Omega/(\pi G\Sigma)$, is $\sim$6 at $r=r_{0}$.  We also vary
$\Sigma_{0}$ from 2$\times10^{-4}$ to 0.01 (g0p2 to g10), and the
corresponding Q varies from 160 to 3.  The effects of the indirect
force and self-gravity become more important in more massive
discs. For runs which only include the indirect force, we add ``i'' at
the end of their model names. For runs which only include disc
self-gravity, we add ``g'' at the end of their names. For runs
including both the indirect force and self-gravity, we add ``gi'' at
the end of their names. For three high resolution runs having 1536
radial grids and 2048 azimuthal grids, the model names are ended with
``H''.  All the runs are summarized in Table 1.

\begin{table*}
\begin{center}
\caption{Simulations\label{tab1}}
\begin{tabular}{cccccccccc}
\hline
\hline
Run   & $\Sigma_{0}$  & Self-gravity & Indirect Force & $R_{c}^{\,a}$ & $\Sigma_{max}(R=R_{c})^{\,b}$ & $\Sigma_{min}(R=R_{c})^{\,b}$ &$\Delta R/R_{c}^{\,c}$\\
\hline
g5  & 5$\times$10$^{-3}$ & No  & No & 0.89  & 8.8$\times$10$^{-3}$  & 3.8$\times$10$^{-3}$ & 0.23\\
g5i  & 5$\times$10$^{-3}$ &    No  & Yes & 0.78 & 1.3$\times$10$^{-2}$  & 1.8$\times$10$^{-3}$ &  0.30\\
g5g  & 5$\times$10$^{-3}$  & Yes  & No  &   0.98 & 6.5$\times$10$^{-3}$  & 5.0$\times$10$^{-3}$ & 0.12\\
g5gi & 5$\times$10$^{-3}$  & Yes  & Yes &   1.03 &  6.5$\times$10$^{-3}$  & 4.8$\times$10$^{-3}$ & 0.15\\
g5giH & 5$\times$10$^{-3}$  & Yes  & Yes &   1.02 & 6.6$\times$10$^{-3}$  & 4.8$\times$10$^{-3}$  & 0.15\\
g0p2i  & 2$\times$10$^{-4}$ & No  & Yes &   0.88 & 3.7$\times$10$^{-4}$  & 1.4$\times$10$^{-4}$  & 0.24 \\
g0p2g  & 2$\times$10$^{-4}$ & Yes  & No &  0.92 & 3.4$\times$10$^{-4}$  & 1.6$\times$10$^{-4}$  & 0.20 \\
g0p2gi & 2$\times$10$^{-4}$  & Yes  & Yes &  0.90 &   3.4$\times$10$^{-4}$  & 1.6$\times$10$^{-4}$ &  0.21 \\
g0p5i  & 5$\times$10$^{-4}$ & No  & Yes &   0.88 &   9.1$\times$10$^{-4}$  & 3.8$\times$10$^{-4}$ &0.22\\
g0p5g  & 5$\times$10$^{-4}$ & Yes  & No &  0.93 &   8.3$\times$10$^{-4}$  & 4.3$\times$10$^{-4}$  &0.20 \\
g0p5gi & 5$\times$10$^{-4}$  & Yes  & Yes &  0.92 &   8.3$\times$10$^{-4}$  & 4.1$\times$10$^{-4}$  &0.20 \\
g1i  & 1$\times$10$^{-3}$  & No  & Yes &   0.87  &   1.9$\times$10$^{-3}$  & 6.6$\times$10$^{-4}$ &0.25\\
g1g  & 1$\times$10$^{-3}$  & Yes  & No &  0.93 &   1.6$\times$10$^{-3}$  & 8.9$\times$10$^{-4}$ &0.18\\
g1gi & 1$\times$10$^{-3}$  & Yes  & Yes &  0.94 &   1.6$\times$10$^{-3}$  & 8.6$\times$10$^{-4}$ &0.19\\
g1giH  & 1$\times$10$^{-3}$  & Yes  & Yes &  0.95 & 1.6$\times$10$^{-3}$  & 8.6$\times$10$^{-4}$  &0.19\\
g2i  & 2$\times$10$^{-3}$ & No  & Yes  &  0.86 &   3.8$\times$10$^{-3}$  & 1.3$\times$10$^{-3}$  &0.25\\
g2g  & 2$\times$10$^{-3}$  & Yes  & No &  0.94 &  3.0$\times$10$^{-3}$  & 1.9$\times$10$^{-3}$ &0.17\\
g2gi & 2$\times$10$^{-3}$ & Yes  & Yes &  0.96 &   3.1$\times$10$^{-3}$  & 1.8$\times$10$^{-3}$ &0.17\\
g10i  & 10$\times$10$^{-3}$ & No  & Yes &  0.51 & 5.8$\times$10$^{-2}$  & 4.3$\times$10$^{-3}$  &0.26\\
g10g  & 10$\times$10$^{-3}$ & Yes  & No &  1.02 & 1.1$\times$10$^{-2}$  & 9.4$\times$10$^{-3}$    &0.15\\
g10gi & 10$\times$10$^{-3}$  & Yes  & Yes &   1.03 & 1.2$\times$10$^{-2}$  & 9.5$\times$10$^{-3}$   &0.15 \\
g10giH & 10$\times$10$^{-3}$  & Yes  & Yes & 1.02 & 1.1$\times$10$^{-2}$  & 9.6$\times$10$^{-3}$   &0.15 \\
\hline
\multicolumn{5}{l}{Various disc $\alpha$}\\
\hline
g5($\alpha$=$10^{-5}$)   & 5$\times$10$^{-3}$  & No  & No &  0.90  & 8.6$\times$10$^{-3}$  & 3.8$\times$10$^{-3}$ &0.22\\
g5($\alpha$=$10^{-4}$)   & 5$\times$10$^{-3}$   & No  & No &  1.00  & 5.1$\times$10$^{-3}$  & 4.8$\times$10$^{-3}$ & 0.09\\
g5($\alpha$=$10^{-3}$)   & 5$\times$10$^{-3}$  & No  & No &   0.88 & 3.7$\times$10$^{-3}$  & 3.7$\times$10$^{-3}$ &-\\
\hline
\end{tabular}
\end{center}
{$^{a}$ $R_{c}$ is the radial position where the maximum surface density lies at 150 orbits.}\\
{$^{b}$ $\Sigma_{max}(R=R_{c})$ and $\Sigma_{min}(R=R_{c})$ are the
  maximum and minimum density along the azimuthal direction at
  $R=R_{c}$ at 150 orbits.}\\
{$^{c}$ $\Delta$ R is the width of the vortex in the radial direction. It is the distance between two radial positions 
where $\Sigma=(\Sigma_{max}+\Sigma_{min})/2$ at the azimuthal angle of the vortex center. }
\end{table*}

\section{Results}

\begin{figure*}
\centering
\includegraphics[trim=0cm 0.0cm 0cm 0.cm, width=1.\textwidth]{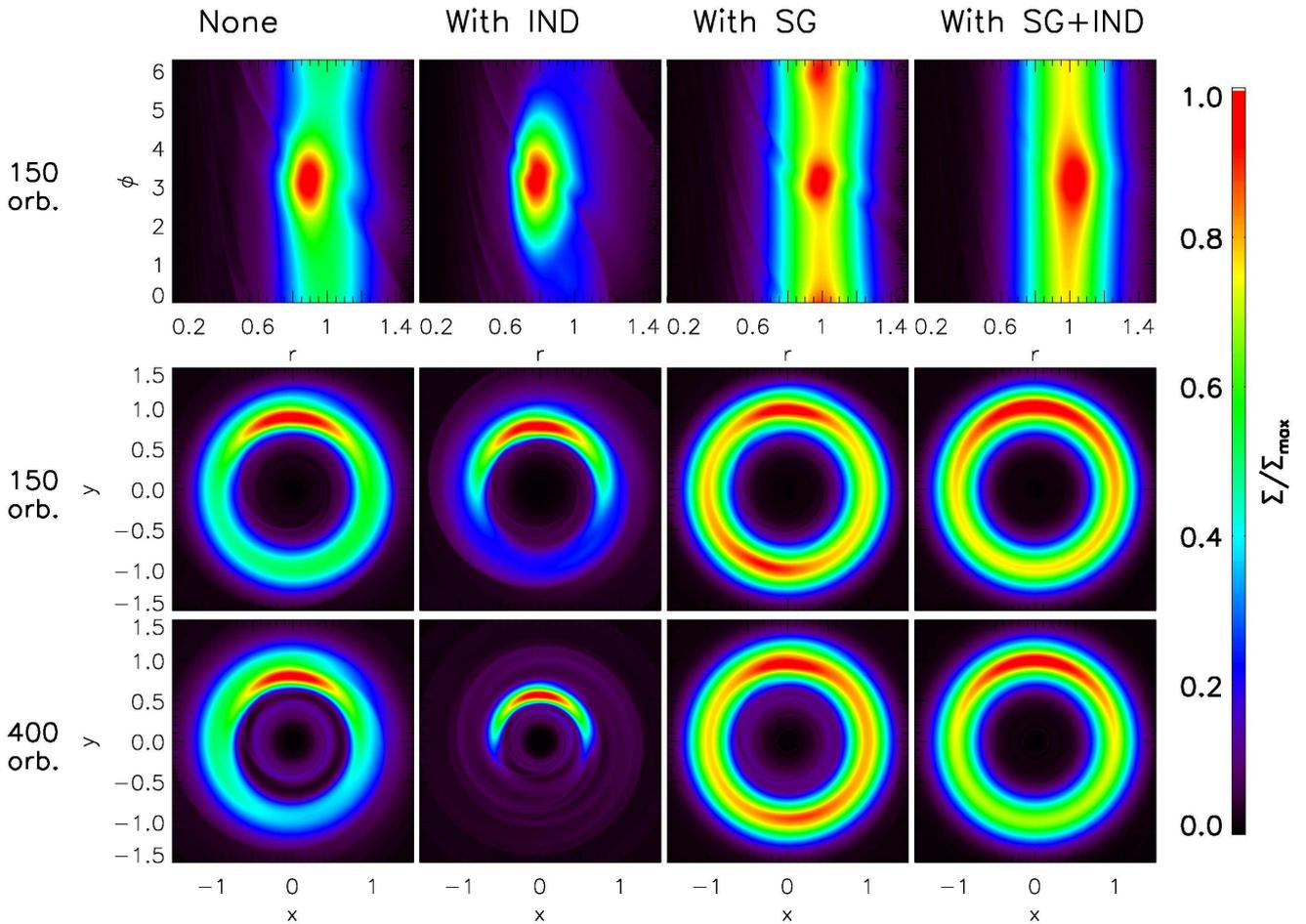} 
\caption{Upper panels: The disc surface density contours at 150 orbits
  for case g5, g5i, g5g and g5gi in the $r$-$\phi$ plane. Middle and
  Bottom panels: The disc surface density contours for the same cases
  at 150 (middle panels) and 400 (bottom panels) orbits in the $x$-$y$
  plane.  } \label{fig:veg}
\end{figure*}

After the simulations start, an $m=1$ mode grows exponentially, and in
most runs it saturates at $\sim$40 orbits. For runs where self-gravity
is important (e.g. g10g), it saturates at a slightly longer time
(e.g. $\sim$60 orbits).  With our fiducial disc mass (g5), the disc
surface density for runs having either the indirect force or disc
self-gravity or both are shown in Figure \ref{fig:veg}.  In Figure
\ref{fig:veg} and all figures below, we have rotate the images so that
the maximum gas surface density is always at $\phi$=$\pi$.
Table 1 gives the radial position where the maximum surface  density lies at 150 orbits, labeled as $R_{c}$.  Since $R_{c}=1$ initially, the deviation of $R_{c}$ from 1 reflects the radial migration speed of the vortex.  At 150 orbits, the maximum and minimum density along the azimuthal direction at $R=R_{c}$ are also  given in Table 1, labeled as $\Sigma_{max}$ and $\Sigma_{min}$. The  ratio between $\Sigma_{max}$ and $\Sigma_{min}$ represents the  strength of the vortex. The vortex radial width based on the vortex density structure is also given
in Table 1.

Figure \ref{fig:veg} clearly shows that the indirect force  strengthens the vortex.
 Compared with case g5 (the leftmost
panels), the vortex in case g5i (the second panel from the left) is
more roundish with a radial width larger than 2 disk scale height, 
and has a higher contrast between the vortex center and
the background (also shown in Table 1). This stronger vortex also
excites stronger density waves. These density waves carry the angular
momentum of the vortex away and enables vortex migration
\citep{PLP10}. With strong density waves excited, the vortex in g5i
migrates inward fast. The vortex in case g5i migrates from $r=1$ to
0.5 within 400 orbits.

{ On the other hand, disc self-gravity stabilizes the vortex.} With
only self-gravity included (case g5g, the third panel from the left),
the saturated $m=1$ mode breaks into two vortices at $\sim$100 orbits
and these two vortices are found to remain separated even at 400
orbits. This is consistent with \cite{GN88}, \cite{LinMK12}, and
\cite{yellin-Bergovoy2015} that self-gravity inhibits Rossby-wave
instability, especially for low $m$ modes.  Quantitatively, linear
instability analysis by \cite{lovelace2013} has suggested that disc
self-gravity can suppress RWI modes with $m<(\pi/2)(r/H)Q^{-1}$.  The
suppression of $m=1$ mode in the case g5g whose
$(\pi/2)(r/H)Q^{-1}=2.5$ is consistent with this criterion.

{ However, when both disc self-gravity and the indirect force are
  included (case g5gi, the rightmost panel), the $m=1$ mode persists
  in the simulation.}  The shift of power from the $m=2$ mode to the
$m=1$ mode when the star is allowed to move has first been seen in
numerical simulations by \cite{christodoulou1992}.  Although disc
self-gravity tries to stabilize the vortex and inhibits the $m=1$
mode, the indirect force tries to maintain the $m=1$ mode.
Eventually, the vortex in case g5gi is weaker than the vortex in case
g5, and it migrates considerably slower than the vortex in either g5
or g5i. Detailed analyses suggest that it migrates less than 0.5\% in
the radial direction within 500 orbits.

\begin{figure*}
\centering
\includegraphics[trim=0cm 0.0cm 0cm 0.cm, width=1.\textwidth]{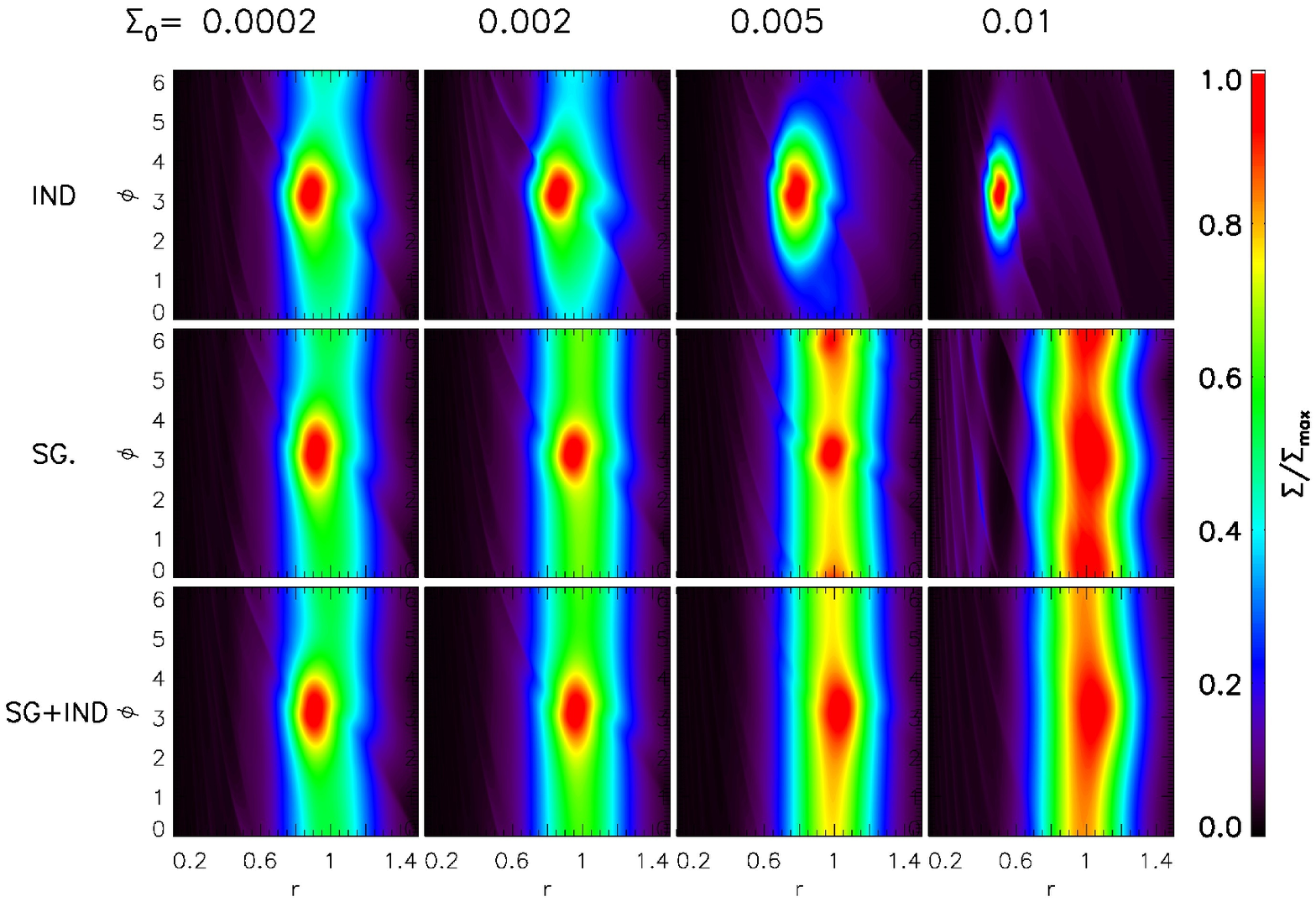} 
\caption{Upper panels: The disc surface density contours at 150 orbits
  for discs having different masses (from left to right panels) if the
  indirect force or self-gravity or both are included (from upper to
  lower panels).  From left to right panels, $(\pi/2)(r/H)Q^{-1}=$0.1,
  1, 2.5, 5, and $m=1$ mode is suppressed when $(\pi/2)(r/H)Q^{-1}>1$.
} \label{fig:vortexvmg}
\end{figure*}

{ Both the indirect force and disc self-gravity become more important
  for shaping the vortex structure when the disc becomes more
  massive.}  Figure \ref{fig:vortexvmg} shows the surface density of
discs having different masses (increasing mass from left to right
panels) if the indirect force or self-gravity or both are included
(from upper to lower panels).  For the smallest disc mass (g0p2), both
disc self-gravity and the indirect force are weak, and they have
negligible effects on the vortex structure. With the disc surface
density increasing, the indirect force starts to strengthen the vortex
and the vortex migrates faster in a more massive disc (upper panels).
On the other hand, disc self-gravity starts to weaken the vortex (the
middle row of Figure \ref{fig:vortexvmg}) in massive discs. As
mentioned above, RWI modes with $m<(\pi/2)(r/H)Q^{-1}$
\citep{lovelace2013} will be affected by disc self-gravity.  From left
to right panels in Figure \ref{fig:vortexvmg},
$(\pi/2)(r/H)Q^{-1}=$0.1, 1, 2.5, 5 respectively.  As expected, $m=1$
mode is completely suppressed in the rightmost two cases, and only
weak $m=2$ perturbations have been observed. In the most massive case,
even the $m=2$ perturbation is significantly suppressed.  When both
the indirect force and self-gravity are included (lower panels), these
two effects seem to be competing with each other, but eventually the
effect of self-gravity seems to be dominant and the vortices become
weaker than those in the top panels.  The indirect force which has
$m=1$ symmetry also leaves its imprint by maintaining the $m=1$ disc
asymmetry.  The middle and bottom panels suggest that, as long as disc
self-gravity is included, the vortex migrates slowly in the disc.

\begin{figure}
\centering
\includegraphics[width=0.5\textwidth]{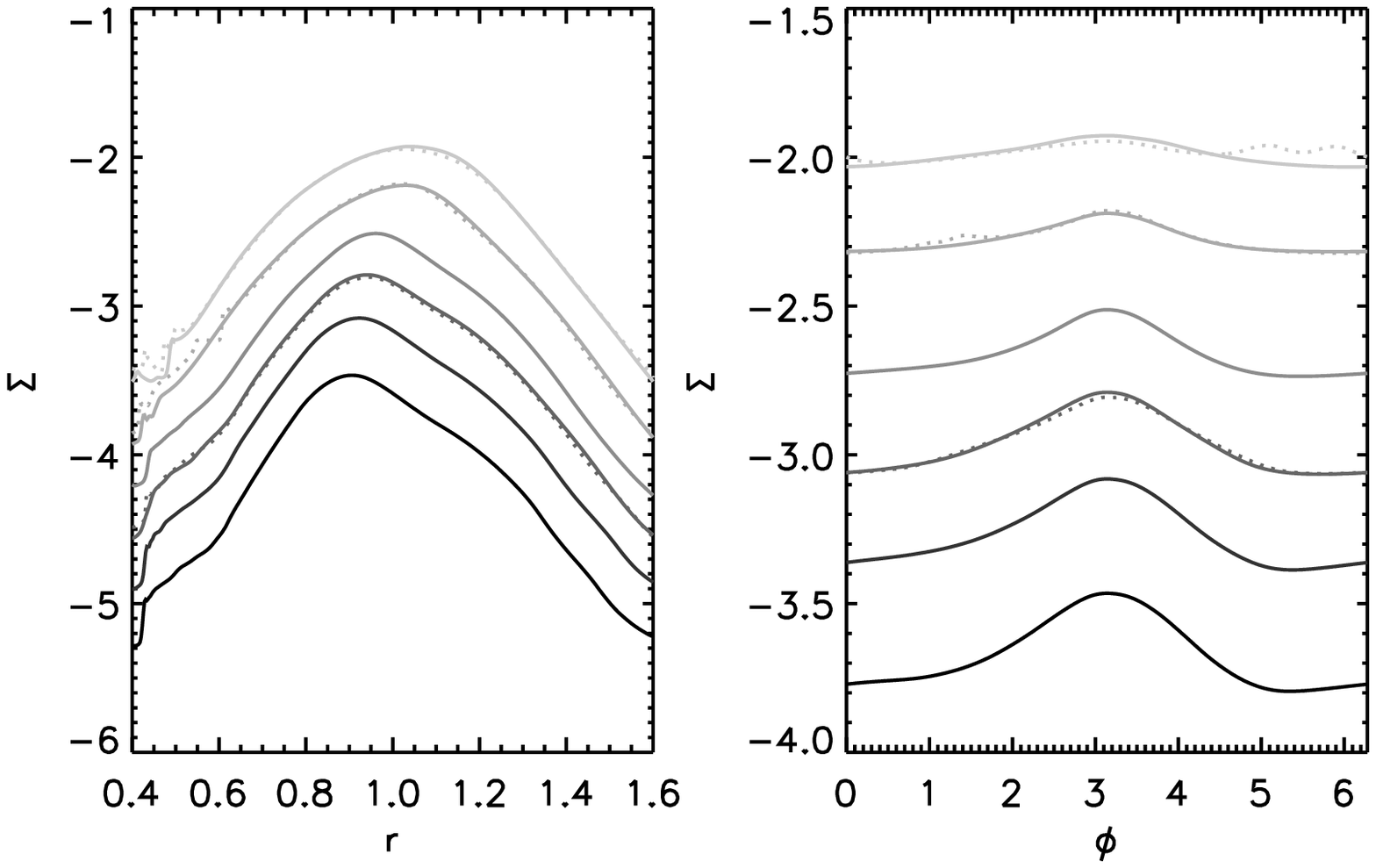} 
\vspace{-0.3 cm}
\caption{Density cuts across the vortex center along the $r$ (the left
  panel) and $\phi$ (the right panel) directions for g0p2gi, g0p5gi,
  g1gi, g2gi, g5gi, and g10gi (from heavy to light curves). The dotted
  curves are from high resolution runs g1giH, g5giH, and g10giH.
} \label{fig:onedcut}
\end{figure}

To quantitatively compare the asymmetric structure in simulations with
different masses, cuts of the disc's surface density across the vortex
center along the $r$ and $\phi$ directions are shown in Figure
\ref{fig:onedcut}. The peak density in the radial profile is at a
larger position for a more massive disc, suggesting that the vortex
migrates inward slower (or even migrates slightly outward) in a more
massive disc.  The density profile in the azimuthal direction clearly
shows that the vortex becomes weaker in a more massive disc due to the
stabilizing effect of disc self-gravity.  The density ratio between
the vortex center and the background is $\sim$2 for the least massive
disc while it is $\sim$ 1.2 for the most massive case. The similarity
between low resolution and high resolution runs suggests that the
simulations are numerically converged. There is a slight difference
for the azimuthal density profile for the most massive cases between
g10gi and g10giH. In g10giH, some small vortices are present which do
not merge with the big vortex at 150 orbits. However, the big vortex
has similar structures as that in the lower resolution run.

 Since the vortex structure sensitively depends on the disc
  viscosity \citep{de2007, Fu2014, zhustone2014}, we have carried out
  three additional simulations with $\alpha=10^{-5}$, $10^{-4}$, and
  $10^{-3}$ using the same setup as our fiducial case g5 which has
  $\alpha=10^{-6}$. Table 1 shows that in the case with
  $\alpha=10^{-5}$ the vortex has a similar strength as our fiducial
  case, but it becomes significantly weaker in cases with bigger
  $\alpha$.  Intuitively, we would expect that the vortex cannot be
  generated when the growth timescale of RWI is comparable with the
  viscous spreading timescale of the density bump ($\Delta
  R^{2}/\nu$).  In our simulations, the width of the density bump is
  $\Delta R =0.2 H$, and the vortex saturates at around 50
  orbits. Thus, $\alpha=0.01$ should suppress the vortex formation. In
  our simulations, a comparable but smaller $\alpha$ (
  $\alpha=10^{-3}$) has completely suppressed the vortex.

\begin{figure*}
\centering
\includegraphics[trim=0cm 0.0cm 0cm 0.cm, width=1.0\textwidth]{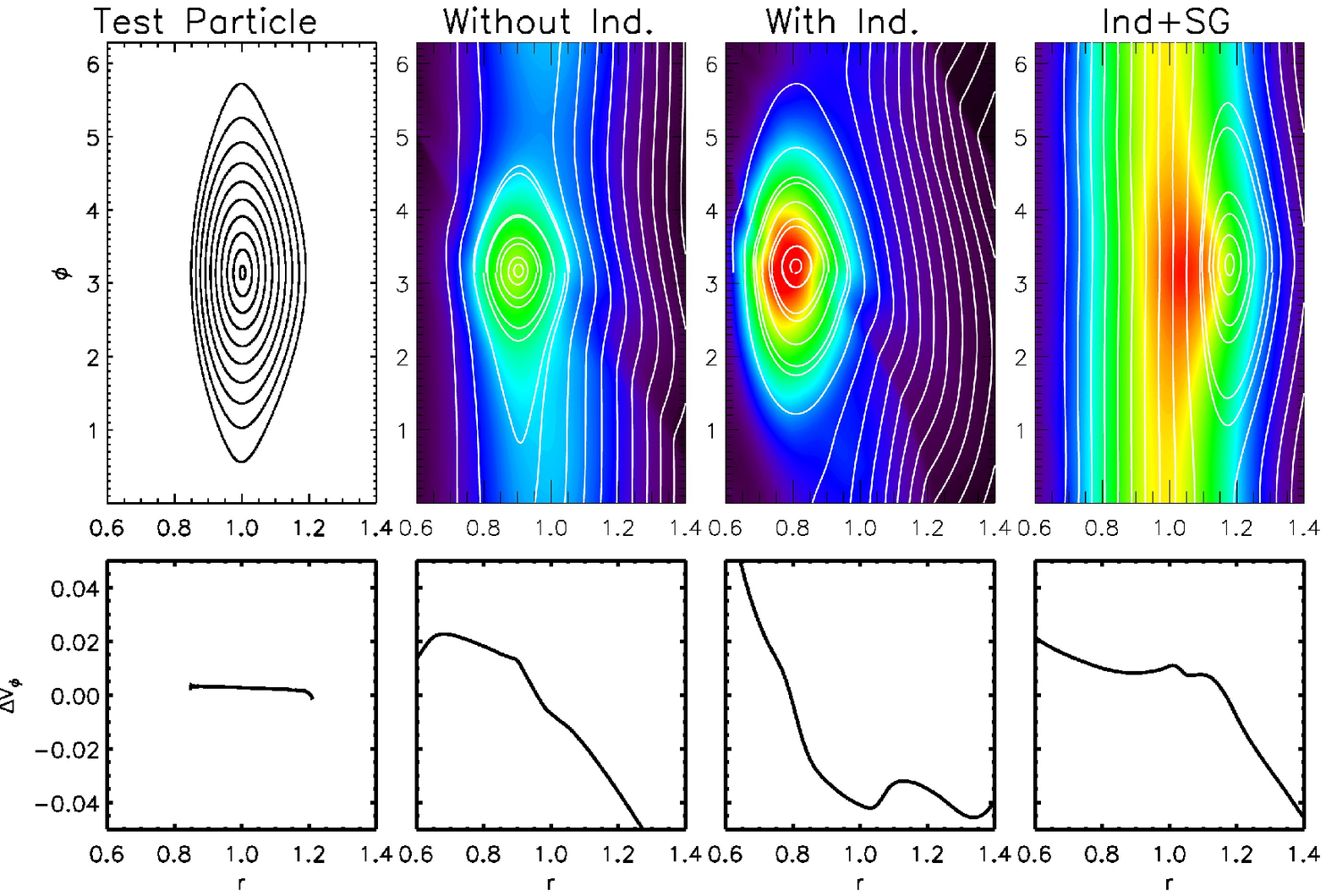} 
\vspace{-0.3 cm}
\caption{The upper left panel: the streamlines in the rotating frame
  for horseshoe orbits in the pressureless fluid (the same as the
  right panel of Figure \ref{fig:indirect}).  Upper middle and right
  panels: the density contours and streamlines in the frame corotating
  with the lopsided structure for g5, g5i, and g10gi at 150 orbits.
  Lower panels: the difference between the azimuthal velocity of the
  fluid element at $\phi=\pi$ and the local Keplerian
  velocity. } \label{fig:stream}
\end{figure*}

To understand how the gas pressure, the indirect force, and disc
self-gravity affect the vortex structure, the density contours and the
fluid streamlines in the frame corotating with the asymmetric
structure are shown in the upper panels of Figure \ref{fig:stream}.
The indirect force alone in the leftmost panel can generate a lopsided
structure which has a similar aspect ratio as the vortex in case g5
(the second panel from the left).  However, they have very different
velocity structures as shown in the bottom panels of Figure
\ref{fig:stream}, indicating the presence of the gas pressure has
significantly altered the fluid velocity structure.  When both the
indirect force and the gas pressure are considered, the indirect force
gives an additional force to spin the vortex, making the vortex
stronger with a smaller aspect ratio (case g5i, the third panel from
the left).  The strong vortex in g5i has a large radial extend.
The radial width measured from the vortex density structure 
is $\sim$3.4 H($R=R_{c}$) (Table 1). The radial width measured from
the largest elliptical vortex streamline in Figure  \ref{fig:stream} is $\sim$4 H($R=R_{c}$).
These results suggest that the vortex can have an envelop extending 
beyond the sonic point of the vortex by a factor of $\sim$2 
(also in Zhu \etal 2014 and Fu \etal 2014).

\begin{figure}
\centering
\includegraphics[trim=2cm 0.0cm 0cm 0.cm, width=0.45\textwidth]{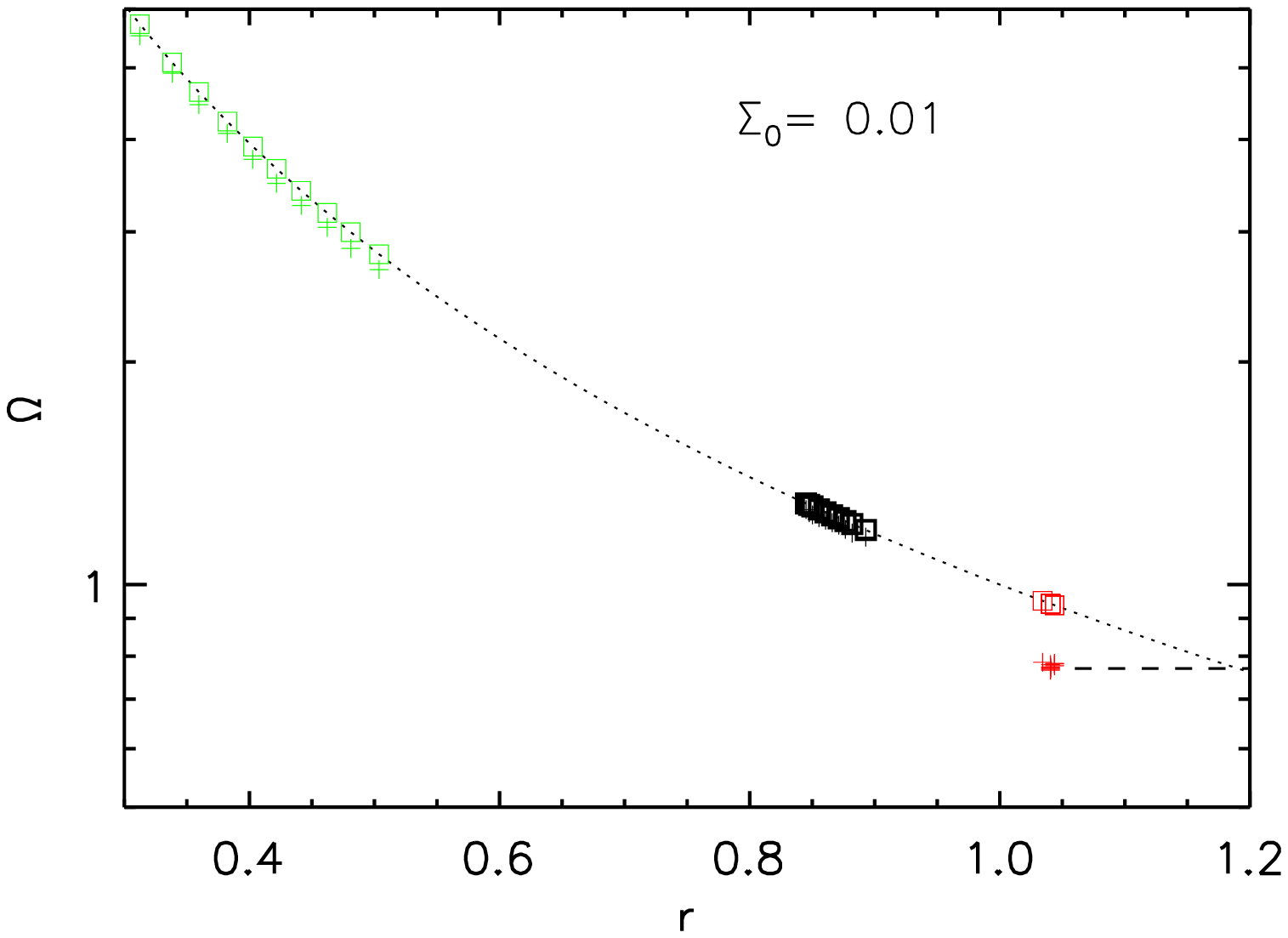} 
\vspace{-0.3 cm}
\caption{The pattern speed with respect to the vortex position in case
  g5 (black pluses), g10i (green pluses), and g10gi (red pluses).  The
  dotted curve represents the Keplerian speed. The squares are
  Keplerian speed at the same radial positions as the pluses.  The
  dashed line labels the pattern speed for the g10gi case. The
  corotation radius is at $r\sim$1.2, further than the vortex itself.
} \label{fig:vortexcenter}
\end{figure}

However, when disc self-gravity is included and the disk is relatively
massive (the rightmost panels in Figure \ref{fig:stream} for case
g10gi), the streamlines remarkably go across the asymmetric structure
without any rotary vortex motion around the density maximum.  The
vortex motion occurs at a much larger distance around $r\sim$1.2, and
the density contours do not coincide with the velocity
streamlines. This is due to the fact that the lopsided structure
orbits around the star at a speed significantly slower than the local
Keplerian speed when the disc self-gravity starts to dominate the
vortex dynamics. In these massive disks, the ``fast mode'' appears to
be transiting to the ``slow mode''.  As shown in Figure
\ref{fig:vortexcenter}, the red pluses suggest that the pattern speed
for the lopsided structure is 80\% of the Keplerian speed in case
g10gi.  The corotation radius where the disc's Keplerian angular speed
\footnote{Strictly speaking, the disc does not rotate at the exact
  Keplerian speed due to disc self-gravity.} matches the structure's
pattern speed is thus at $r\sim$1.2 instead of $r\sim$1. In the frame
corotating with the disk asymmetric structure as shown in the
rightmost panel of Figure \ref{fig:stream} , the zero velocity point
is thus at $r\sim$1.2.

 We would expect that such sub-Keplerian asymmetric disk
  structure cannot trap dust particles at the gas maximum.  This is
  because dust particles always try to move at the local Keplerian
  speed and they will move in and out of the sub-Keplerian asymmetric
  structure quickly and cannot be trapped. But for particles at the
  corotation radius, they will remain at the same relative position
  with the asymmetric structure and can potentially be affected by the
  asymmetric disk structure.  When we include dust particles in the
  simulations (Paper II), some particles indeed concentrate to the
  corotation radius at $r\sim$1.2.  Particles outside the vortex
  motion at $r\sim$1.2 feel a non-zero gas drag and sink to the center
  of this vortex motion.   This interesting radial offset between the
gas and dust lopsided structures will be shown in Paper II and may be
observable with ALMA.

The detailed force balance for the vortex in various cases has been
presented in Appendix A, where we can see the pressure gradient plays
an important role in determining the vortex structure.

\section{Conclusions}
We have studied asymmetric flow patterns in massive discs.  We have
first searched possible ``fast modes'' (lopsided structures orbiting
around the central star at the Keplerian speed) in pressureless fluid
using test particle methods. Without the indirect force that comes
from the displacement of the central star by the lopsided structure,
the only ``fast mode'' is the traditional pressureless anticyclonic
vortex with an aspect ratio ($\chi$) of 2. When the indirect force is
considered, besides the pressureless vortex mode, another ``fast
mode'' whose streamlines consist of slowly librating horseshoe orbits
exist, as shown by \cite{MC15}.  The streamlines in this mode are
highly elongated with very large aspect ratios.

When the gas pressure is included, these two modes are less distinct.
Together with the indirect force, the gas pressure significantly
alters the lopsided structure.  Vortices with any aspect ratio are now
possible even due to the gas pressure alone \citep{goodman1987}.
Vortex with a larger aspect ratio rotates around the vortex center
slower, and can be more easily affected by other additional forces
(such as the indirect force and self-gravity).

Using two dimensional global hydrodynamical simulations with both the
indirect force and disc self-gravity self-consistently included, we
have studied the effect of the indirect force on the vortex structure,
which is initiated by the Rossby Wave Instability.  The simulations
confirm that the indirect force alone can widen the vortex streamlines
to be more than two disc scale heights wide in the radial direction
when the mass ratio between the disc and the central star is
$\gtrsim$1$\%$, and the vortex migrates faster when the indirect force
becomes important.  However, for such discs, disc self-gravity becomes
equally (if not more) important than the indirect force.  In massive
discs where $(\pi/2)(r/H)Q^{-1}>1$ \citep{lovelace2013}, disc
self-gravity alone suppresses the $m=1$ mode. But the indirect force
restores the $m=1$ lopsided structure.  Still, overall, the vortex is
significantly weakened by disc self-gravity. Vortices' inward
migration slows down. Vortices in some cases can even migrate
outwards.

One important observation is that, when the disc is massive enough
(e.g. Q$\sim$3), the lopsided gas structure orbits around the star at
a speed significantly slower than the local Keplerian speed.  In this
case, there is a radial shift between the lopsided structure itself
and the corotation radius where the disc's Keplerian rotation matches
the structure's pattern speed.  Since dust can be trapped at the
corotation radius, it suggests that there could be a radial shift
between the gas and dust distributions in vortices of massive discs.
Dust distribution in vortices in massive discs are presented in Paper
II.  Overall, disc self-gravity is important to regulate the vortex
structure in massive discs and it has observational signatures which
may be probed by current and future observations.

Our simulations have several limitations. First, we have only
  explored a small parameter space of disc structures.    
  Besides
  $H/r=0.1$, we have also explored discs with $H/r=$ 0.05, 0.07, 0.08,
  and 0.09 at $r_{0}$ while keeping $\sigma=2 H(r_{0})$ in Equation
  \ref{eq:sigma0}.  Significant growth of the RWI has been observed
  only when $H/r>\sim0.09$ despite the fact that the density bump is always
  2  scale height wide in each case, implying the vortex production by the RWI
  is sensitive to the disc and density bump structure. Second, our 2-D
  simulations cannot capture vortex instabilities occurring in 3-D,
  such as the elliptical instability \citep{LP2009}.  The effects of
  indirect force and disc self-gravity on these instabilities need
  further study. Moreover, in massive discs where the gas and dust
  lopsided structures have a radial offset, how such offset affects
  the parametric instability from dust to gas feedback also needs to
  be revisited \citep{RP2014}.  

\section*{Acknowledgments}
All hydrodynamical simulations are carried out using computer
supported by the Princeton Institute of Computational Science and
Engineering, and the Texas Advanced Computing Center (TACC) at The
University of Texas at Austin through XSEDE grant
TG-AST130002. Z.Z. acknowledges support by NASA through Hubble
Fellowship grant HST-HF-51333.01-A awarded by the Space Telescope
Science Institute, which is operated by the Association of
Universities for Research in Astronomy, Inc., for NASA, under contract
NAS 5-26555.

\appendix
\section{Force Balance within The Vortex}
\begin{figure*}
\centering
\includegraphics[trim=0cm 0.0cm 0cm 0.cm, width=1.0\textwidth]{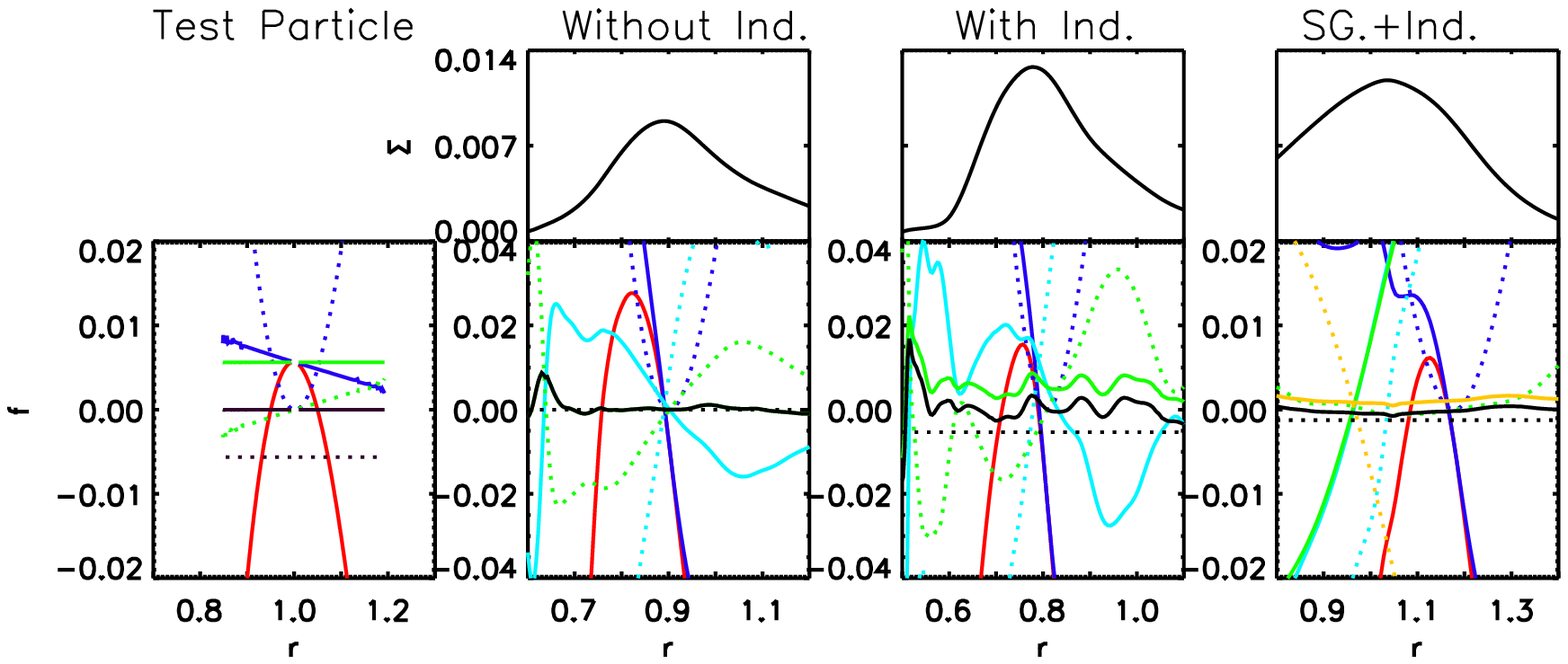} 
\vspace{-5.2 cm}
\caption{The disc surface density and force balance with respect to
  $r$ at $\phi=\pi$ for horseshoe orbits in pressureless fluid, g5,
  g5i and g10gi at 150 orbits (from left to right).  Bottom panels:
  various force components.  The red curve is $-\frac{\partial
    \Phi_{*}}{\partial r}+\Omega_{f}^{2}r+2\Omega_{f} v_{\phi}$, the
  blue dotted curve is $\frac{v_{\phi}^2}{r}$, the cyan dotted curve
  is $-\frac{\partial P}{\rho \partial r}$, the green dotted curve is
  $-v_{r}\frac{\partial v_{r}}{\partial r}-v_{\phi}\frac{\partial
    v_{r}}{r\partial \phi}$, the black dotted curve is
  $-\frac{\partial \Phi_{ind}}{\partial r}$, and the orange dotted
  curve is $-\frac{\partial \Phi_{sg}}{\partial r}$. The solid curves
  are the total forces which add one additional force represented by
  the dotted curve in the same color.  } \label{fig:paperr}
\end{figure*}

The force balance at $\theta=\pi$ for various cases is shown in Figure
\ref{fig:paperr}. For a circular orbit, the tidal force $-\partial
\Phi_{*}/\partial r+\Omega_{f}^2 r$, Coriolis force (2$\Omega_{f}
v_{\phi}$) and centrifugal force ($v_{\phi}^2/r$) balance each other.
When only the indirect force is included (the left most panel), the
velocity at $r=1$ becomes slightly super-Keplerian since the Coriolis
force and the centrifugal force need to balance the indirect force.
Thus, the velocity in the rotating frame at $r=1$ becomes
$v_{\phi}\sim v_{K}+f_{ind}/2\Omega$ where $v_{K}$ is the Keplerian
velocity.  The leftmost panel of Figure \ref{fig:paperr} also suggests
that the force which turns the vortex ($-v_{r}\frac{\partial
  v_{r}}{\partial r}-v_{\phi}\frac{\partial v_{r}}{r\partial \phi}$)
is at a similar amplitude as the indirect force. With the gas pressure
included (right panels), the pressure gradient (the cyan dotted curve)
is quite large, and its balance with other forces leads to the vortex
rotation.  When disc self-gravity has been included (the rightmost
panels), the self-gravitational force becomes zero at $r\sim0.95$.
The pressure gradient becomes zero at $r\sim 1$. To balance the
self-gravitational force at $r\sim 1$, the addition of tidal force,
Coriolis force, and centrifugal force has to be larger than zero at
$r\sim 1$. This is possible if the corotation radius is at $r>1$.

\bibliographystyle{mnras}
%\bibliography{paper}

\label{lastpage}
\end{document}